\documentstyle[prb,aps]{revtex}
\newcommand{\be}{\begin{equation}}
\newcommand{\ee}{\end{equation}}
\newcommand{\beq}{\begin{eqnarray}}
\newcommand{\eeq}{\end{eqnarray}}

\def\o{\omega}

\def\s{\sigma}
\def\p{\partial}
\def\h{\hbar}
\def\r{\rho}

\def\e{\varepsilon}

\begin{document}
\title{ Search for tunnelling centres in Lennard-Jones clusters
	}
\author{ 
	G.~Daldoss, O.~Pilla and G.~Viliani
	}
\address{ 
	  Dipartimento di Fisica and Istituto Nazionale di Fisica della 
	  Materia, \\ Universita' di Trento, I-38050 Povo, Trento, Italy
	  }
\maketitle
\date{\today}
\begin{abstract}
We report on numerical procedures for, and preliminary results on the search
for, tunnelling centres in Lennard-Jones clusters, seen as simple model
systems of glasses. Several of the double-well potentials identified are
good candidates to give rise to two-level systems. The role of boundary
effects, and the application of the semiclassical WKB approximation in
multidimensional spaces for the calculation of the ground state splitting
are discussed.   
\end{abstract}
\section{Introduction}
The thermal properties of amorphous solids at low temperature ($T \approx
0.1 \div 10$ K) are qualitatively different from those of the corresponding
crystals (Zeller and Pohl 1971). The observed temperature dependence of the
heat capacity ($C_v(T) \propto T^3$) and of the thermal conductivity
($\kappa(T) \propto T^{2 \div 3}$) of crystalline solids at temperatures
much lower than the Debye temperature is well understood in terms of a
density of vibrational states  (which is the only factor determining $C_v$)
$\r(\o) \propto \o^2$, and by assuming that phonons carry heat in the same
way
as particles do in a gas. On the contrary, in a large variety of amorphous
or disordered solids (covalent glasses, polymers, disordered crystals, spin
glasses, etc) the corresponding observed behaviour is $C_v(T) \propto T$,
$\kappa (T) \propto T^2 $; moreover, the heat capacity is 2-3 orders of
magnitude larger than in crystals, while the contrary is true for the
thermal conductivity. This behaviour is so widespread and independent on
structural details that any reliable explanation or model should be based on
very general properties of the disordered state. 
\par
The idea that since 1972 has gained more-or-less general consensus is the
so-called model of the Two-Level Systems (TLS) (Anderson, Halperin and Varma
1972; Phillips 1972, 1987). In the disordered structure the local mass
distribution is rather dishomogeneous and may produce holes of different
sizes; in this situation it is possible that atoms, or groups of atoms, may
have two (or more) equilibrium positions available, separated by energy
barriers. The
energy separation between the two lowest levels in the two-well system
depends on the energy mismatch of the two minima and on the tunnelling
splitting. If the minima are nearly at the same energy, it is the tunnelling
splitting that determines the energy difference between the two lowest
levels, and depending on the potential shape this splitting can be very
small.
This situation produces additional possibilities of low-energy thermal
excitation with respect to crystals (and so $C_v$ increases) but, on the
other hand, low-frequency, heat-carrying phonons will be scattered by the
TLS
by either  direct absorption or  relaxational phenomena (and thus $\kappa$
decreases).
\par
Since both the mismatch and the tunnelling splitting depend on the local
details of the disordered structure, it is reasonable to assume (Anderson et
al 1972, Phillips 1972) that the distribution of TLS excitations is
constant, at least for small splitting, resulting in a linear $C_v(T)$ at
low $T$ (for a derivation of both $C_v(T)$ and $\kappa(T)$ see the review
paper by Phillips (1987)). 
\par
The microscopic nature of the TLS can only be investigated numerically
because disorder prevents analytical solutions of the vibrational 
problem. The efforts produced so far in this direction (see for example
Weber and Stillinger 1985, and references therein; Heuer and Silbey 1993,
1997, and references therein; Demichelis, Ruocco and Viliani 1997) have
employed molecular dynamics simulations on samples containing from the order
of 100 to the order of 1000 atoms, with periodic boundary conditions
imposed. This approach has advantages and disadvantages. 
\par
One advantage is that molecular dynamics can nowadays handle rather large
systems, and the use of periodic boundary conditions eliminates surface
effects. One disadvantage is that in any case 1000 atoms means a simulation
box having a size of 10 atoms, and with these dimensions periodic
boundary conditions can introduce spurious periodicity and
correlations. Moreover (and maybe more important), finding the
configurations corresponding to two potential energy minima is not enough to
calculate the splitting or even to decide whether the pair is a suitable
candidate to be a TLS, because the splitting depends on the shape of the
potential barrier that separates the minima; for example, in the case of a
single particle of mass $m$ moving along a 1-dimensional trajectory, the
semiclassical WKB approximation (Froman and Froman 1965; Landau and Lifchitz
1967) gives the
following expression for the tunnelling splitting in a symmetric two-well
potential $V(x)$:
%%% 1
\be
	\delta= \frac{\h\o}{\pi} D^{\frac{1}{2}}
\ee
Here the transmission coefficient $D$ is given by
$$
	D=\frac{1}{1+\text{exp}(2S_0)},
$$
and $S_0$ is the action integral:
$$
	S_0=\frac{1}{\h} \int_{-a}^a \sqrt{2m(V(x) -E_0)} dx.
$$
In the last formula $E_0$ is the particle energy in one well, $a$ and $-a$
are the classical
turning points, and $\o$ is the oscillation frequency in one well. 
\par
Equation (1) provides an explicit expression for the splitting in terms of
the potential and as such is very convenient for computations, but
its extension
to multidimensional configuration spaces is not obvious in general.
The difficulty stems from the fact that in multidimensional spaces
(infinitely) many different paths
contribute to the dynamics, each one yielding a probability amplitude $p_i$
for the tunnelling event whose total probability is then given by
$$
P=|\sum_i p_i |^2
$$
One should actually use the formalism of Feynman path integrals
(Feynman and Hibbs 1965; Gillan 1987; Voth, Chandler and Miller 1989;
Ranfagni, Mugnai, Moretti and Cetica 1990; Schenter, Messina and Garret
1993) to evaluate the
probability. However in many circumstances the main contribution to the
probability comes from the {\it least action path} (Ranfagni and Viliani
1976; Ranfagni, Viliani, Cetica and Molesini 1977; Ranfagni et al 1990), and
the multidimensional problem turns into a 1-dimensional one. The conditions
under which this simplification can be made
are discussed at length by Ranfagni et al (1990); basically they depend on
the possibility that, and on the extent to which, the wave equation can be
(approximately) separated into different
equations, each involving a single independent variable (Schiff 1968).
\par
The accomplishment of this approximate separation is also important in the
classical case, where the barrier is overcome by thermal activation. The
one-dimensional transition rate is formally very similar to equation (1):
$$
	k=\frac{\o_0}{2\pi}\; \text{exp}(-\frac{E_b}{k_B T})
$$
where $E_b$ is the barrier energy and $\o_0$ the vibration frequency in the
potential well. In many dimensions, and under some assumptions concerning
the existence of thermodynamic equilibrium and the absence of back-
crossings, it is found that the degrees of freedom other than the single
considered path introduce entropy barriers, and that the effect of the
latter can be accounted for by the following substitution in the pre-
exponential factor (Rice 1958; Glyde 1967; Hanggi 1986):
$$
  \o_0 \rightarrow      \frac{\Pi_i\; \o^M_i}{\Pi'_j\; \o^S_j}.
$$
Here the $\o^M$'s and the $\o^S$'s are the vibrational eigenfrequencies at
the minimum and at the saddle point respectively, and the prime indicates
that the negative frequency at the saddle point has to be omitted from the
product. We shall assume that this substitution takes proper account of
entropic effects also in the tunnelling case, and shall use the WKB formula
(1) along the least action path to evaluate the splitting.
\par
From the above considerations,
it appears that knowledge of the saddle points (or "transition states" in
the chemist's jargon) is of paramount importance for the study of
tunnelling or diffusion problems because a good
initial guess for the least action path is one that connects the minima
through the saddle point itself. The methods of molecular
dynamics alone are not especially designed for their identification: finding
the minima is
relatively easy by means of a variety of efficient numerical methods
(repeated quenching and/or viscous forces in molecular dynamics, conjugate
gradients,
simulated annealing, and so on), but the saddle points are much harder. As a
consequence of this state of affairs, previous works were able to identify
only a limited number of possible TLS. 
\par
The approach of the present paper is complementary to the traditional ones
in two respects. 
\par
First of all, we consider free Ar clusters rather than systems with boundary
conditions. We are aware that surface effects will be serious, but we think
it is very important to have information regarding both surface-free and
correlation-free systems. Furthermore, Ar clusters are interesting
on their own and by increasing the number of the constituent atoms it should
in principle be possible to make the cluster properties coalesce with those
of the bulky solid (Buck and Krohne 1994). So our plan is to study the
evolution of tunnelling-related properties as the number of atoms in the
cluster increases.
\par
Second, for the reasons listed above, our search is primarily directed to
finding large numbers of
saddle points in addition to minima. As mentioned, the methods devised by
Weber and Stillinger (1985),
Heuer and Silbey (1993, 1997) and Demichelis et al (1997) are not specific
to this task. So we will look for all kinds of
stationary points on the potential hypersurface by finding the zeros of the
gradient. As will be discussed in the next section, in the case of the
Lennard-Jones potential this requires the solution of a non linear set of
equations. Alternative methods for systematic saddle point search are
described in the review paper of Berry (1993), where numerous references to
work on minima  are also listed. 
\par 
%%%%%%
\section{Numerical procedures}
\subsection{Stationary configurations of Ar${_N}$ clusters}
%%%%%%
We consider Ar${_N}$ clusters in the range $N=6 \div 42$; the
atoms interact via a Lennard-Jones pair potential and the total potential
energy is
%%% 2
\be
	V(x_1,...,x_{3N})= \sum_{ij}V_{ij}=
		4\e \sum_{i>j} \big[\large(\frac{\s}{r_{ij}})^{12} -
(\frac{\s}{r_{ij}})^6]
\ee
where $\s=3.405 \AA$ and $\e/K_B=125.2 K$ are the parameter values suitable
for argon. 
\par
In a free non-rigid body like a Lennard-Jones cluster there are
problems with rotations. In fact, in a rigid body these (like translations)
are decoupled from the vibrational degrees of freedom, and correspond to
zero-frequency eigenvalues, but this is no longer true for a deformable
body, because the Coriolis and centrifugal forces couple vibrations and
rotations. The problem is negligible in the majority of minima
because in these configurations the restoring forces are in general strong
enough to quench the coupling, but it becomes severe in the
saddle points: these are unstable configurations and the coupling to
rotations can produce very large effects. On the other hand, if we have in
mind clusters of atoms which are part of a
solid, i.e. the set of atoms that undergo the largest displacements when a
tunnelling transition occurs, then these clusters certainly are not allowed
to rotate or translate. For this reason we have decided to eliminate 6
degrees of freedom
by fixing the position of atom 1 in the origin ($x_1=y_1=z_1=0$), while atom
2 is bound to the $x$ axis ($y_2=z_2=0$) and atom 3 can move on the $xy$
plane ($z_3=0$). These conditions were used also by Hoare (1979) and do not
influence the potential energy of the
configurations, which depends solely on the mutual distances of the atoms,
but definitely affect the vibrational frequencies: clusters that have the
same
shape (i.e. are placed in the same minimum of the configuration space) but
different atoms with fixed
coordinates, have also different vibrational eigenvalues. This effect
becomes
less and less important as $N$ increases and, on the other hand, since the
fixed coordinates are chosen randomly we do not expect that our results may
be biased in any systematic way.
\par
Let us now summarize the procedure by which we find the stationary
points of Eq. (2). Writing down the partial derivatives of $V$ is tedious
but
straightforward; once they are known, the non linear set of equations:
%%% 3
\be
	\frac{\p V(x_1,...,x_{3N})}{\p x_i} =0, \;\;\;\;\;\; i=1,...,3N-6
\ee
is solved by using the Newton-Raphson (NR) iterative procedure (Press,
Flannery, Teukolsky and Vetterling 1986). A random configuration of atoms is
created and taken as starting point for the first iteration step; the
calculation requires also knowledge of the second partial derivatives. With
a given initial random configuration, the procedure may or may not converge
to a solution. If convergence is not achieved in a given number of
iterations (in
our case 5000), another initial configuration is created. When the
procedure converges (typically in our case after a few thousand steps) the
Hessian $\cal{H}$ is diagonalized in the final configuration and the number
of
negative eigenvalues gives the order of the saddle point (zero-order
corresponds to a minimum, while maxima are not possible for this potential).
The configuration, its energy and the Hessian
eigenvalues are stored and the procedure is started again with a new random
configuration. In the case $N<=13$ the program has been run until no new
stationary point was found for a few ten hours of a 160 MHz Pentium cpu
time, or for a few hours of a Digital Alpha 2100 cpu time. 
\par
As noted in previous works (Hoare 1979, Weber and Stillinger 1985) the
saddle points that matter really are the first order ones, because between
two neighbouring minima in configuration space there always is at least a
first-order saddle point, and saddles of higher order lie on average at
higher energy. So, especially for larger clusters we limited the search to
first-order saddles (this actually means that only first order saddle points
and minima were stored, because there is no known way of preferentially
ending in either of these kinds of configurations with the NR procedure).
The number of saddle points is not known but is certainly extremely large
and grows enormously with $N$. Thus for $N>13$ we made no attempt at
completeness and limited ourselves to collecting some thousand first-order
saddles of different energy for each $N$. 
\par
The direct search for first-order saddles, however, works well only for $N<
7$ because for larger clusters
the number of higher-order (and higher energy) stationary points is so
huge that the NR procedure very often ends up in one of them. This
happens also because the initial random cluster usually has a very high
potential energy, i.e. it is very distant from any of the desired
configurations, and this is a well-known wrong start for NR. In order to
solve this problem, before starting the NR procedure the initial random
configuration is relaxed towards a minimum either
by the conjugate gradient method or by a molecular dynamics calculation
including a viscous force. Both algorithms are very efficient in reaching
the vicinity of a minimum, that we label $M_1$. Once such approximate
minimum is
found, we repeatedly diagonalize the Hessian and move upwards in energy in
the direction of the eigenvector corresponding to the minimum eigenvalue:
this path leads us towards the saddle point. When the maximum potential
energy along this path is reached, the system is in general sufficiently
near the saddle point
that the NR procedure locates it with great precision in a small number of
iterations. In the majority of cases this approach singles out a first order
saddle, but in a limited number of instances it may end up in a higher
order one, in a minimum, or even fail to converge.
%%%%%%%%%
\subsection{Least action path}
%%%%%%%%%%
Once the saddle point is located with precision, in order to calculate the
classical action integral we move the representative point away from the
saddle in the direction of the negative eigenvalue, and let it evolve either
by molecular dynamics plus viscous force, or following at each step the
direction of minimum eigenvalue. This is done from both sides of the saddle,
and in this way we identify the two minima $M_2$ and $M_3$ that are
connected by the saddle in question. It is possible that neither $M_2$ nor
$M_3$ coincide with $M_1$. 
\par
The resulting path is often rather close to the least action path and in
some cases further minimization of the action integral according to the
method described by Demichelis et al (1997) produces only minor changes.
In other cases, the straight path from one minimum to the other is
closer to the least action path.
%%%%%%%%%
\section{Preliminary numerical results}
%%%%%%%%%
\subsection{Stationary configurations}
%%%%%%%%%
In Fig. 1 we report a semilogarithmic plot of the number of minima $g(N)$ as
a
function of the number of atoms up to $N=13$. If one disregards the cases
$N=2\div 6$ which have only 1 or 2 minima each and cannot be taken into
account for statistical considerations, $g(n)$ grows exponentially
%%% 4
\be
g(N)=A\;\text{exp} (bN), \,\,N>6
\ee
with $A=(3.1 \pm 0.8) \times 10^{-3}$ and $b=0.99 \pm 0.03$. This is a
good check that the large majority of minima have been found since $g$
determines the extension of the configuration space where there exist stable
states, $\Gamma$, and this in turn determines the entropy of the system
according to the relationship $S=K_B \text{ln} \Gamma$: an exponential $g$
thus produces an extensive entropy. The more-than-exponential growth found
by Hoare (1979) is due to his consideration of the small clusters ($N<7$) in
the fitting procedure. 
\par
From Eq. (4) we obtain $g(15) \approx 10,000$ and $g(18) \approx 200,000$,
so that, as mentioned, we didn't even try to determine all minima for
$N>13$, but limited ourselves to finding some thousand of them for clusters
of increasing size. For $N=15,\;18$ this was done in two different ways: (i)
starting from a random initial configuration; (ii) relaxing the clusters
from the first-order saddle points found in the previous step. The
distributions of minima energies obtained in
these two cases for $N=15$ are reported in Figs. (2a) and (2b) respectively;
as can be seen the distributions are practically identical, indicating that
we introduce no special bias by starting the search for tunnelling centres
from the first-order saddle points. Similar results are obtained for $N=18$.
This is a very
important point because this approach greatly facilitates the identification
of the centres with double-well potentials (DWP) and their characterisation.
\par 
Fig. 3 shows the distribution in energy of 16,875 minima and 9,176 first-
order saddle points for $N=29$. The distributions look very similar, with
the obvious difference that the saddles are on average at higher energy. 
%%%%%%%%
\subsection{Double-well potentials}
%%%%%%%%%
The pairs of minima separated by a first order saddle have been identified
with the procedure described in the previous section. The results are
summarized in Table I, where we report the total number of DWP identified,
the number of symmetric ($\Delta =0$) and of slightly asymmetric ($0<\Delta
<1 \;K$) DWP, and the number of DWP that have both small asymmetry and small
tunnelling splitting ($10^{-15}\; K< \delta < 1\;K$). 
The DWP in the last 
column are the suitable ones to produce two-level systems; their number
relative to the total DWP ranges from 0.9$\times 10^{-3}$ for $N=13$, to
3.8$\times 10^{-3}$ for $N=42$, and apparently grows with $N$, though the 
numbers are too small to extract reliable trends.  It should also be
 noted that the values in
the last column of Table I were obtained by taking $E_0=\frac{1}{2}\h\o_0 $
in the calculation of the action integral, where $\o_0$ is the minimum
eigenvalue of the dynamical matrix in the minimum. This procedure 
is right only if the
direction of the least action path is the same as that of the lowest-energy
eigenvector, and in general it overestimates the action integral. Therefore,
these values are to be considered as lower bounds.  
\par
As regards surface effects, we find that the vast majority of the symmetric
($\Delta=0$) DWP involve
large surface motions as indicated by the huge action integrals and the long
euclidean distance; as can be seen from Table I their relative number
decreases steadily with increasing $N$ as it should; however, the abrupt 
fall for $N=29,\;42$ is probably due to biased search.
\par
Another interesting characteristic of the DWP is the so-called
participation number, defined as
$$
{\cal N}= \sum_i d_i^2/d_{M}^2 = d^2/d_{M}^2
$$
where $d$ is the euclidean distance between the two minima in configuration
space and $d_M$ is the displacement of the atom that moves most. This
quantity gives an indication of how many atoms move significantly when the
system passes from one minimum to the other, and the results of its
calculation in the cases $N=18,29,42$ are reported in Fig. 4 for all DWP,
and in Fig. 5 for selected DWP with asymmetry $0<\Delta <1 \; K$, i.e. the
possible TLS. It is interesting to note that, in both figures, the maximum
number of occurrences seems to have nearly reached saturation at $N=42$.
Moreover, there seem to be no large qualitative differences between the
corresponding distributions of Fig. 4 and of Fig. 5: it appears that the
participation number does not depend very much on the asymmetry. A similar 
analysis based on the value of the barrier height gives analogous results,
in the sense
that DWP that are candidates to be TLS have an ${\cal N}$ distribution not
qualitatively different from the whole assembly of DWP. The present results,
yielding an average value of about 15 participating atoms, are in general
agreement with previous values found with periodic boundary conditions
(Heuer and Silbey 1993).
%%%%%%%%%%
\section{Conclusions and perspectives}
%%%%%%%%%%
In this paper we have described a set of algorithms that are very efficient
in finding minima and saddle points of multidimensional (potential-energy)
surfaces. In particular, we have studied Lennard-Jones clusters containing
up to 42 atoms and were able to find many thousand minima and
first-order saddle points; among these we looked for possible  
two-level-systems and found several probable candidates in clusters   
with more than 6 atoms (see Table I), i.e. pairs of minima with total 
splitting in the ground state smaller than $\approx 1 \;K$.
\par
To evaluate the ground state splitting one must consider both the asymmetry
of the pair of minima, $\Delta$, and the tunnelling splitting, $\delta$. 
While $\Delta$ is easily found, the calculation of the tunnelling splitting
requires that the Schroedinger equation is solved in a way or another. We
chose the semiclassical WKB approach; this has the advantage of providing
an explicit expression for $\delta$, but at the same time its application
to multidimensional problems requires great caution (see previous 
discussion). 
\par
Another point that deserves further consideration is the role of boundary
conditions in the simulation of the properties of {\it small} systems. 
Clusters suffer from surface effects, while periodic boundary 
conditions are likely to introduce spurious correlations. This may
explain the large difference in the number of minima, DWP and candidate 
TLS between this
work and the paper of Heuer and Silbey (1993). 
\par
Since for large enough systems boundary conditions must become
irrelevant, it would be worth  to check if for 
periodic systems the rate of growth of, for example, the total number
of minima tends to approach the behaviour of Fig. 1; for this, it
would be very interesting to compute (possibly) all the minima of 
periodic systems. 
\par
Finally, we plan to apply the present analysis to potentials suitable
to experimentally available glasses.
\par
{\bf Acknowledgments.} We wish to thank G. Ruocco, L. Angelani and 
G. Parisi for very useful discussions and suggestions.

%%%%%%%%%%%%%%%%

\newpage
FIGURE CAPTIONS\\
\vskip .7cm
\noindent
Fig. 1. Semilogarithmic plot of the number of minima, $g$, as a function of
the
number of atoms. The best fit gives $g(N) \propto \text{exp}(N)$ (see
text).
\vskip .7cm
\noindent
Fig. 2. Energy distribution of minima for $N=13$ obtained starting from:
(a) an initial random configuration, (b) first-order saddle points.
\vskip .7cm
\noindent
Fig. 3. Energy distribution of 16,785 minima (open circles) and 9,175 first-
order saddle points (full squares) for $N=29$.
\noindent
\vskip .7cm
\noindent
Fig. 4. Participation number of all DWP for $N=18$ (a), $N=29$ (b), $N=42$
(c).
\vskip .7cm
\noindent
Fig. 5. Same as Fig. 4, but only for DWP with asymmetry $0<\Delta < 1\; K$.

\vskip 1cm
\noindent
\narrowtext
\begin{table*}
\caption
{Number of found DWP for $N=6 \div 42$.\\
 (a): DWP with $\Delta=0$; 
(b): $10^{-4}<\Delta<1\;K$;
(c): Same as (b) but with barrier higher than minimum vibrational eigenvalue
and $10^{-15}< \delta < 1 \;K$. }
\begin{tabular}{ccccc} 
N& Total No. of DWP &(a)&(b)&(c)\\
 \hline
6&6&5&0&0\\
\hline
8&61&26&0&0\\
\hline
9&181&58&4&0\\
\hline
10&414&113&2&0\\
\hline
13&3416&588&41&3\\
\hline
15&4652&257&43&4\\
\hline
18&11412&1082&150&19\\
\hline
29&9176&23&63&18\\
\hline
42&2878&12&31&11\\
\end{tabular}
\end{table*}

\newpage
\widetext
\begin{center}
REFERENCES\\
\end{center}
\vskip 1cm
\noindent
Anderson P.W., Halperin B.I., and Varma C.M., 1972, Philos. Mag. 25, 1.\\
Berry R.S., 1993, Chem. Rev. 93, 2379.\\
Buck U., and Krohne R., 1994, Phys. Rev. Lett. 73, 947.\\
Cozzini S, 1994, PhD Thesis, Universit\`a di Padova, unpublished.\\
Cozzini S., and Ronchetti M, 1996, Phys. Rev. B53, 12040.\\
Demichelis F., 1996, Thesis, Universit\`a di Trento, unpublished.\\
Demichelis F.,  Ruocco G., and Viliani G., 1997, to be published.\\
Feynman R.P., and Hibbs A.R., 1965, {\it Quantum Mechanics and
Path Integrals} (MacGraw-Hill, New York).\\
Froman N., and Froman O.O., 1965, {\it JWKB Approximation} (North-Holland,
Amsterdam).\\
Gillan M.J., 1987, J. Phys. C20, 3621.\\
Glyde H. R., 1967, Rev. Mod. Phys. 39, 373.\\
Hanggi P., 1986, J. Stat. Phys. 42, 105.\\
Heuer A., and Silbey R.J., 1993, Phys. Rev. Lett. 70, 3911.\\
Heuer A., and Silbey R.J., 1996, Phys. Rev. B53, 609.\\
Hoare M.R., 1979, Adv. Chem. Phys. 40, 49; and references therein.\\
Landau L., and Lifichitz E., 1967, {\it M\'ecanique Quantique}, chapter
7 (MIR, Moscow).\\
Phillips W.A., 1972, J. Low Temp. Phys. 7, 351.\\
Phillips W.A., 1987, Rep. Progr. Phys. 50, 1657.\\
Press W.H., Flannery B.P., Teukolsky S.A., and Vetterling W.T., 1986,
{\it Numerical Recipes} (Cambridge University Press, Cambridge).\\
Ranfagni A., and Viliani G., 1976, Solid State Commun. 20, 1005.\\
Ranfagni A., Viliani G., Cetica M., and Molesini G., 1977, Phys. Rev. B16,
890.\\
Ranfagni A., Mugnai D., Moretti P., and Cetica M., 1990, {\it Trajectories
and Rays: the Path Summation in Quantum Mechanics and Optics} (World
Scientific, Singapore).\\
Rice S.A., 1958, Phys. Rev. 112, 804.\\
Schenter G.K., Messina M., and Garrett B.C., 1993, J. Chem. Phys. 99,1674.\\
Schiff L.J., 1968, {\it Quantum Mechanics} (McGraw-Hill, New York).\\
Voth G.A., Chandler D, and Miller W.M., 1989, J. Chem. Phys. 91, 7749.\\
Weber T.A., and Stillinger F.H., 1985, Phys. Rev. B32, 5402.\\
Zeller R.C., and Pohl R.O., 1971, Phys. Rev. B4, 2029.\\

\end{document}